# A Review of Routing Protocols for Mobile Cognitive Radio Ad Hoc Networks

S. Selvakanmani[1] and Dr. M. Sumathi[2]

[1]Assistant Professor, Department of Computer Science and Engineering, Velammal Institute of Technology, Chennai, India
sskanmani6@yahoo.com

[2]Professor, Department of Electronics and Communication Engineering, Velammal Engineering College, Chennai, India
nnsuma@yahoo.com

## ABSTRACT

*Ad hoc network is a collection of wireless mobile nodes that dynamically form a temporary network without the use of any existing network infrastructure or centralized administration. A cognitive radio is a radio that can change its transmitter parameters based on interaction with the environment in which it operates. The basic idea of cognitive radio networks is that the unlicensed devices (cognitive radio users or secondary users) need to vacate the spectrum band once the licensed device (primary user) is detected. Cognitive capability and reconfigurability are the key characteristics of cognitive radio. Routing is an important issue in Mobile Cognitive Radio Ad Hoc Networks (MCRAHNs). In this paper, a survey of routing protocols for mobile cognitive radio ad networks is discussed.*

## 1. INTRODUCTION

Cognitive radio is an emerging technology, which is used to solve the problem of scarce spectrum resource utilization. Ad hoc networks are self – organizing and adaptive, i.e. it can take different forms [1], whereas the cognitive radio networks can change its transmitter parameters according to the interactions with the environment in which it operates [2, 13]. Cognitive network can be defined as an intelligent network since it automatically senses the environment and current network conditions, and adapts the necessary communication parameters accordingly.

In Cognitive radio network, two forms of users exist. They are the primary user and the secondary user. Primary users (PUs) have high priority than the Secondary Users (SUs) in the utilization of the spectrum [6]. The Cognitive radio [2] enables the unlicensed (secondary) users to sense the unoccupied spectrum portions which are not used by the licensed (primary) users for a specific amount of time. There are two forms of cognitive radio networks. They are 1. Primary networks and 2. Cognitive radio (CR) networks. The following section describes the classification in detail.

## 1.1 Classification of Cognitive radio networks

The Classification of Mobile Cognitive radio Ad hoc networks [4] is of two types, which is described in the following Figure 1.

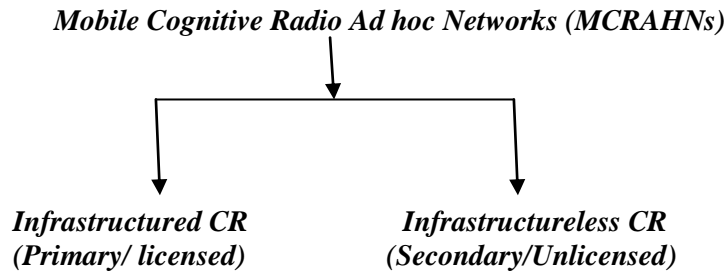

Figure 1. Classification of CR networks

In *infrastructured networks,* a central, fixed infrastructure component called base station will be present for the communication among the communicating devices. In this network [3], the primary users have licenses to operate in certain spectrum bands. The primary user [PU] activities are controlled through primary base stations. The components of the primary network are: Primary User (PU), Primary base station.

In *infrastructureless networks,* the communication among the devices is performed without the support of the fixed component. The Unlicensed network [3] does not have a license to operate in a desired band. The main component of the secondary network is the Secondary User (SU). Figure 2 shows the difference between the two networks in their corresponding licensed and unlicensed band, in the form of network architecture.

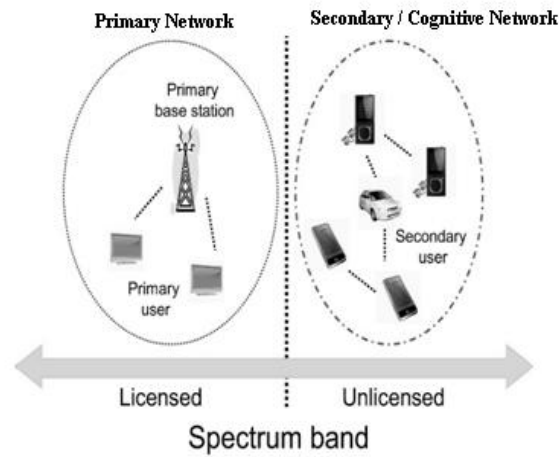

Figure 2. CR network architecture (in both licensed and unlicensed band)

## 1.2 Cognitive Radio and Ad Hoc networks

Both the Cognitive Radio and Ad Hoc networks are different in nature. They differ in terms of (1) Transmission Spectrum Opportunities (SOP) (2) Multi Spectrum transmission [5]. Moreover, in ad hoc networks, two nodes communicate among themselves, if they are in the same channel and frequency i.e. fixed spectrum. And there is no difference in the mobile users.

The parameters used for network connectivity, among them are the distance and the transmission power. In Cognitive Radio Ad hoc networks, the spectrum followed is in dynamic nature. Any two nodes can connect to themselves if they are in different radio visibility and at the least one SOP should be available. The interference of the primary user by the secondary user has to be eliminated. The parameters used for network connectivity are the nodes position, transmission power and the SOP.

The other difference between the ad hoc networks and the Cognitive radio ad hoc networks is in terms of mobility of the nodes. In Ad hoc networks, the routes formed over multiple hops may periodically get disconnected due to the mobility of the node. In CR ad hoc networks, a node may not be able to transmit immediately if it detects the presence of a Primary user on the spectrum. [6]

In MCRANs, unidirectional links are more possible[4]. Such links will be available for a fraction of time instead of minutes or hours. This cause's unidirectional links as a failure, as there is no guarantee that a channel used for sending station will be available till the receiving station uses the same channel for transmission.

Due to such dynamic nature of CR links, traditional routing is not easy to maintain the routing table in MCRAHNs. A local on-demand routing or table driven routing is used in reality, at a reasonable routing delay to route packets. The protocols discussed in the paper are the AODV (Ad hoc On Demand Distance Vector) protocol [15] and DSDV protocol (Destination-Sequenced Distance-Vector) [16].

*Spectrum opportunities (SOPs)* are defined as the set of frequency bands currently unoccupied by the Primary Users and therefore, it made available for the Secondary Users [7]. It is also called as *Spectrum Hole or White Space.* These white spaces are the wastage in the RF spectrum and would be secondary user for its communication [4]. The ability of the secondary user to change its frequency is called Dynamic Spectrum access (DSA). [6]

DSA senses the spectrum band and fetches the available opportunities. Thus these networks use the licensed band in opportunistic manner for communication. These communication opportunities are available when primary users have fewer usage of their frequency band or they are not using it at all [4]. Hence availability of channel is time and geographic based. Figure 3 illustrates the Spectrum Opportunities with the Secondary user exploitation. The empty spaces are called white space or spectrum hole.

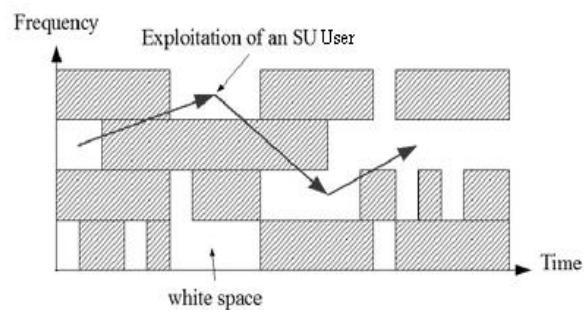

Figure 3. Spectrum Hole or White Space

The rest of the paper is organized in the following manner. Section II comprises about the routing protocols of CR ad hoc networks. Section III concludes the work and the references used for the work are presented.

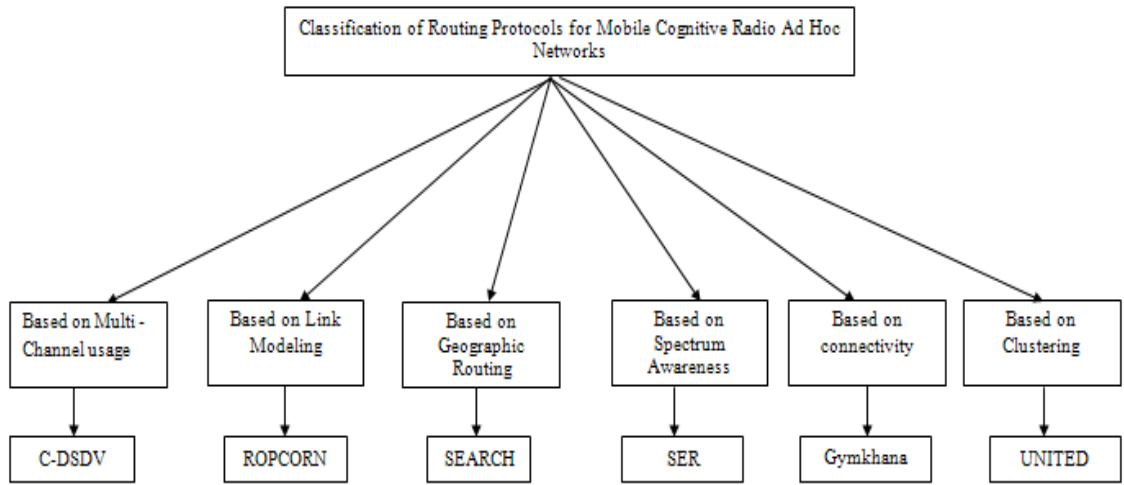

Figure 4. Classification of Routing Protocols for Mobile Cognitive Radio Ad Hoc Networks (MCRAHNs)

## 2. CLASSIFICATION OF ROUTING PROTOCOLS FOR MCRAHNS

In this section, survey of state – of – the – art routing protocols for Mobile Cognitive Radio Ad Hoc Networks are presented. The authors consider the routing protocol for the last three years. The Classification is based on: Multi – channel usage, Link Modelling, Geographic routing, Spectrum awareness, Connectivity and Clustering. The Classification of the routing protocols is presented in the Figure 4.

### 2.1. Based on Channel Switching

Switching of channel frequently occurs when a node wants to change its channel as it has more than two routes to the destination in different channels.

### 2.1.1 C-DSDV Protocol

The protocol is a pre - active cognitive multi – channel routing protocol which utilizes multiple channels in the CR ad hoc networks [8]. The routing and the channel allocation are the two main challenges in CR ad hoc networks, which are considered here in this paper. It aims at optimizing the system performance of multi-hop CR ad hoc networks, by using a cross – layer design approach. The protocol is the enhanced version of the already existing DSDV (Destination Sequenced Distance Vector) protocol [16], for cognitive routing scheme.

Function of the C-DSDV protocol: Each node in the CR ad hoc network maintains a routing table with all possible destinations, and the number of routing hops to each destination is recorded. Therefore the routing information is present every time, even if the source node requires a route or not. Routing table updates are sent periodically throughout the network to maintain the table consistency [8]. New route broadcasts will contain the address of the destination node, the number of hops to reach the destination, the sequence number of the information received about the destination. The sequence number will unique for each broadcast and the route with the updated sequence number is always used for broadcasting.

The nodes will transmit the updates immediately if new information arrives or a change in the topology or a switch in the channel. There is a possible of delaying the route advertisement but the channel change is advertised immediately to all other nodes in the network. The Table 1 summarizes the C – DSDV protocol.

Table 1. Summary of C-DSDV Protocol

| Objective / Features | Algorithm used / Explanation |
|---|---|
| Goal | To optimize the system performance of multi-hop CR ad hoc networks by using multichannel |
| Working of the protocol | The routing table are updated immediately if there is change in the channel |
| Techniques used | Combines the channel allocation and routing |
| Route Discovery | Common control channel used for broadcasting the updation about the routing table |
| Protocol nature | Table Driven Routing Protocol |
| Best Path Selection | Yes. The Sequence number is used for this purpose. The Sequence number which of most recent one is considered as the best path |
| Routing Overhead | Yes. If the Primary users increase, there it is possible to have more routing messages in order to maintain the consistency among the routing tables |
| Metric Used | Change of channel. i.e. Channel Switch |
| Unique Feature | The frequent channel Switch metric improves the system performance |
| Simulator used | Ns2 Simulator |
| Related Protocol | DSDV (Destination Sequence Distance Vector)[16] |

## 2.2 Based in Link Modelling

Links are formed during the communication among the two nodes in a network. The link connectivity and disconnectivity plays an important role in the cost metric.

### 2.2.1 ROPCORN Protocol

The protocol [9] was designed for data transportation by making use of link modelling. Two routing metrics are considered for this purpose. They are spectrum availability cost and load estimation. Aim is to maximize the data rates and minimize delay and the total resources consumed, for a set of communication sessions [9]. The following example reveals the protocol approach.

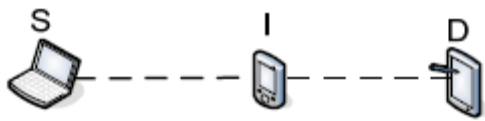

Figure 5. An example for link connectivity/disconnectivity

In the above Figure 5, three nodes are available namely, S, I and D. The S is for the source, I for the intermediate and D for the destination. The link, which is of finite, is available between S – I and I – D. At normal time, S can compute the best path available to D through I. However, if the S finds that the link S – I goes down, it buffers the packet for a fixed short period and forwards to I when the link is up. Similarly, if the link between I – D goes down, and the node I happens to know that, then it will buffer the packet till I – D link comes up. Therefore, ROPCORN protocol routes the packet to the nearest node of the destination or to the network where the destination is connected. [9] Each message of the protocol consists of unique message identifier, a hop count and an optional ack request. Two important characteristics of CR ad hoc networks are considered. They are Spectrum availability and load estimation.

*Spectrum availability*: It is taken as a metric to minimize the latency with minimal buffer usage [9]. The link's connectivity will be tracked by assigning the cost metric, which gets updated periodically to reflect the state. If a link is disconnected for some time, thereby increasing the cost, but for the stable link the cost is kept to a small value. When two nodes have an opportunity to access the spectrum without the interference of the PU, then they form the unidirectional links in the network. The scenario is the same which was explained earlier. A node keeps track of its link by sending "Hello" messages to the neighbors. All the information regarding the neighbors is stored in the neighbor table [9]

*Load estimation:* To measure the traffic load of the CR nodes, the network traffic information is required for the optimization of routing. Since every CR node is aware about the traffic information that crosses itself and its neighbors, then the protocol select a path which is lightly loaded for a new communication session.

*Forwarding method:* Using this protocol, when a message arrives at a node, there might be no path for forwarding due to the PU activity. In such case, the node has to buffer those packets and upon having an opportunity to use spectrum, the decision must be made on whether or not to transfer the packet [9]. Moreover, the same packet has to forwarder to many numbers of nodes in order to increase the probability of delivering the packet to the destination. The Table 2 summarizes the ROPCORN protocol.

## 2.3 Geographic forwarding

The idea is to discover several paths, which are combined at the destination to form the path with the minimum hop count. The nodes used in this approach will be equipped with GPS devices.

### 2.3. 1 SEARCH Protocol - Spectrum Aware Routing protocol

The SEARCH protocol uses the geographic forwarding. This protocol jointly considers the path and the channel selection to avoid the regions of the Primary User activity during the route formation. Minimization of hop count to reach the destination is done by using the optimal path found by geographic forwarding [10]. The idea of the geographic forwarding is used in this protocol. It is able to deal with reasonable levels of PU activity changing rate. Also, a mechanism for disseminating the destination location both at the source and at each intermediate node is required. [11].

The protocol assumes the primary users' activities in an ON/OFF process. The functions followed by the protocol are (1) Route setup phase (2) Joint Channel – Path optimization phase and (3) Route Enhancement, in order to improve the route during its operation. [10]

Table 2 SUMMARY OF ROPCORN PROTOCOL

| Objective / Features | Algorithm used / Explanation |
|---|---|
| Goal | To maximize the data rates and minimize delay for a set of user communication sessions. |
| Working of the protocol | Broadcast the packets through the better link, which is of low cost, without affecting the PU user |
| Techniques used for forwarding | Buffering at the intermediate node and forwards to the correct destination |
| Route Formation | Uses Hello Protocol to know about the neighbors and the link. |
| Protocol nature | On demand routing protocol |
| Best Link Selection | The link with less disconnection is considered as it has lower cost. |
| Routing Overhead | No. Routing overhead is less as the spectrum sensing the beacon messages. |
| Metric Used | Spectrum availability and Load Estimation |
| Unique Feature | Captures any spatial or temporal locality of link disconnection and leveraging it for optimal route selection |
| Simulator used | Ns2 |
| Related Protocol | RACON [3] |

ROUTE SETUP: A route request packet (RREQ) is transmitted by the source to all the nodes of the channel which is not affected by the PU activity. The packet will be forwarded until it reaches the destination, and the intermediate nodes will be adding the following fields along with the RREQ packet. They are: (i) ID, (ii) its current location, (iii) TTL and (iv) flag (to indicate the propagation mode of the algorithm used). The protocol operates in two modes – **Greedy Forwarding and PU Avoidance.**

Greedy Forwarding features are: 1. RREQ will be forwarded on the same channel and the next hop will be not on the PU coverage region. The intermediate node must lie in the specific region around the current hop. If a PU region is interrupted then the present channel can not be used effectively. This is called PU avoidance stage and the RREQ gets surround around the affected region.

If a path exists between the source and the destination and there exists a transmission radius, which centered from the source node and extends up to a maximum angular spread, is called the **focus region** of the source node. However, a node that lies in the focus region of the previous hop along the path, and there is no forwarding node in its own focus region, is called the **decision point.** [10]

JOINT CHANNEL – PATH OPTIMIZATION: This algorithm combines the channels and the propagation paths in order to minimize the end – to – end latency. The steps followed by this algorithm are:

*Step 1: Initial Path Selection*

The destination node receives the RREQs from different channels and extracts the path information from all the RREQs. The path, p, contains a set of nodes with their own timestamps.

*Step 2: Greedy Path Formation.*

In this step, the protocol tries to improve the betterness of the route, by intersecting the paths on the other channels. The decision point are considered for this purpose where takes the optimized path starts and deviates towards those locations. A route which has a shorter route to the destination is considered as the better route, which is from different channels. If the next hop happens to be the final destination, then the algorithm is terminated, the decision follows on Step 5 and the path is found. Else the process will be following the Step 3.

*Step 3: Decision point optimization*

Once the decision point is reached, the protocol finds the intersecting path of different channels on the same location. Example, a given path $P_1$ with the node x is said to be intersecting with another path $P_2$, if it has a common node in common or it has a node within the transmission range of node of $P_1$.

*Step 4: Route Expansion*

The optimal path is updated with the new channel and more information about the path. The final path includes the channel switching decisions and the allowed distance between any two nodes in the network.[10]

*Step 5: Route Confirmation.*

Once the destination is reached, the route reply RREP packet is sent back to the source on the optimal path. RREP includes the identification number of the individual node, their current location, and the channel switching decisions. The data transmission will be started immediately after the reception of the RREP packet from the destination.

ROUTE ENHANCEMENT: The stage which arrives after the initial route setup is the route enhancement stage. It links several paths together formed on different channels that are up to 'n' hops away. The algorithm for enhancing the current path is as follows:

1. From the route setup phase, it is understood that all possible path are checked at each Decision Point (DP). If none of the node is within the transmission range of the source, then the DP is considered as the current path. So, the currently identified shortest path can be further optimized by considering the following (i) all other DPs, (ii) the anchor locations on other paths or channels which with their transmission range.

2. After the discovering of a feasible path by the destination, it sends a Route Enhancement (ROP) messages to the recent DP. The ROP message consists of ID of the recent node which is on the new path, the path information.

3. The DP receiving the ROP sends a RREQ message as a source, to another node which is considered as the destination. The process is similar to the route setup phase involving the greedy finding. The route information received from the ROP is then used for RREP message.

4. Once the destination receives the RREQ, it checks the delay of the path and it has to be less than the current one. If it is less than the current one, the destination node generates the RREP message and forwards to the source. It also generates the RERR message, if the route breaks down during the data transmission.

5. The whole process of run – time optimization is performed until no other optimal path with minimal delay is found.

Following is an example of a combination of route and spectrum discovery. Let's consider a three-dimensional system, with the x-y plane representing the physical space where the CR network and the PUs are located. The z-axis shows the frequency scale and also the different channel bands. The shaded regions in the figure show that a single PU may affect several channels (frequencies) around its location. [14]

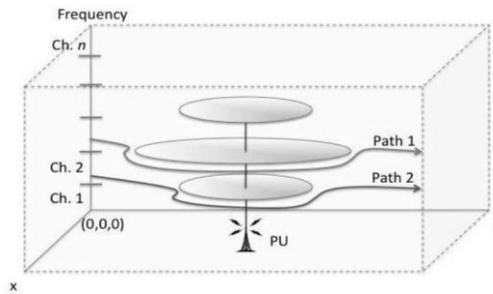

Figure 6. Joint Route and Spectrum Discovery example

Moreover, the channels may be affected to different geographical extents, depending upon their frequency separation with the PU's transmission channel. SEARCH attempts to find paths which circumvent the PU coverage regions (Path 1 and 2) and link them together, whenever a performance benefit is seen. The Table 3 summarizes the SEARCH protocol.

Table 3 SUMMARY OF SEARCH PROTOCOL

| Objective / Features | Algorithm used / Explanation |
|---|---|
| Goal | It undertakes both the path and the selection, to avoid the Primary User activity region. Tries to maintain the end – to – end latency. |
| Working of the protocol | It is based on Geographic routing, which is based on the dynamic spectrum availability and the node mobility. |
| Techniques used for forwarding | Greedy Forwarding and PU Avoidance |
| Route Formation | Based on Geographic Routing |
| Protocol nature | On demand routing protocol |
| Best Path Selection | Yes. It uses the joint channel – path optimization algorithm to find the best path |
| Routing Overhead | Yes. Routing overhead occurs due to the RREQ, RREP, RERR, ROP messages |
| Unique Feature | It's a distributed protocol. Optimizes both the path and the channel |
| Simulator used | Ns2 |
| Related Protocol | Enhanced Version of AODV |

## 2.4 Based on Spectrum Awareness:

The key factor of spectrum aware routing is the combination of spectrum discovery and the route discovery in MCRAHNs.

### 2.4.1  SER - Spectrum and Energy Aware Routing Protocol

The main aim of this protocol to establish a bandwidth guaranteed QoS routes in small CR networks where the topological changes are low. The protocol uses Time Division Multiple access [12]. The QoS requirement considered here is the number of transmission timeslots for a packet on its route from source to reach the destination.

Working of SER protocol: The SER is an on demand routing protocol proposed for multihop CR networks. The basic operation of SER includes route discovery, data transmission and route maintenance.

ROUTE DISCOVERY: The route request (RREQ) broadcast procedure is based on Dynamic Source Routing Protocol (DSR). When the source CR user, say S, has packets to send to the destination CR user, say D, then S will start the route discovery process by broadcasting a spectrum aware RREQ message on the Common Control Channel (CCC). The RREQ message will be received by all of its neighbors. An intermediate CR user, e.g. V, which uses a timer, once it receives the first RREQ for each received RREQ from the neighbor, say U, before the timer expires, V runs the communication segment assignment algorithm to find the feasible communication segments between the link. i.e. $l = (U,V)$. If the feasible communication segment is not found, V drops the corresponding RREQ. Else, the link attaches itself to the current partial path, updates the other information and rebroadcast it. Simultaneously, the Cognitive user increases the value of the hop count and the time to live [12].

The destination user waits for more RREQs to arrive at its side. The destination user also sets up a timer on its side, in order to monitor the multiple RREQs. The destination will use the same technique followed by the source node, to select the optimal path. It computes the route by using the utility route of the path and selects the one which is used to the maximum.

Destination replies the source CR user with route reply (RREP) packet through the same control channel. The communication segment of each link for the path is used especially only for the data transmission, which is a reserved one. When the RREP is forwarded towards the source CR user, all the intermediate users reserves the same communication segment mentioned in the reply packet. There may be problem of intersecting path from other communication segment too. Therefore, instead of finding a new route, the communication segment assignment algorithm coordinates the conflicting users with it s one – hop neighbors scheduling to recover the route reservation.

DATA TRANSMISSION: After receiving the Route Reply by the destination, the data transmission begins. The data will be forwarded by the source to the destination via the intermediate CR users.

ROUTE MAINTENANCE: The route maintenance is re constructed automatically by using route recovery (RREC) and route error (RERR) messages. This phase is important as the users are mobile. The Table 4 summarizes the SER protocol.

### 2.5 Based on Connectivity

It aims at connectivity of the cognitive radio ad hoc networks, based on the movement of the PUs.

### 2.5.1 GYMKHANA Protocol

The protocol that is capable of identifying the network connectivity towards the destination is the Gymkhana protocol. It is a routing protocol that identifies all possible paths of connectivity towards the destination. The protocol forwards the information in the path which is not affected by the network zone, as it does not support the network stability and connectivity. For this purpose, a mathematical framework based on Laplacian spectrum graphs, is used for the evaluation of the different routes of the CR network. [7]

It is a distributed protocol and it is able to measure the connectivity of different network paths and to forward the data packets among the different paths which avoid the network zones. Therefore, by evaluating the activity of the PUs, a path is determined with the highest connectivity. The protocol used three different zones for the differentiation about the connectivity. For example, the authors named a zone called *Red Zone,* which corresponds to the poor network connectivity due to the PUs movement. *Blue Zone* is the zone which is holding all the less movable nodes.

TABLE 4 SUMMARY OF SER PROTOCOL

| Objective / Features | Algorithm used / Explanation |
|---|---|
| Goal | To provide high throughput in multi – hop Cognitive radio ad hoc networks |
| Working of the protocol | It select energy efficient route and assign channels and timeslots for the connection request |
| Techniques used | Communication segment assignment algorithm |
| Route Discovery | Common control channel used for broadcasting the Route Request message |
| Protocol nature | On demand routing protocol |
| Best Path Selection | Yes. It used to increase the lifetime of the CR user individually. Increases the throughput and end – to end latency |
| Routing Overhead | Yes. If the Primary users increase, there it is possible to have more routing overhead due to the RREQ, RREP, RERR, RREC messages |
| Unique Feature | Balances the traffic load among the different CR users |
| Simulator used | An Discrete event simulator written in C language |
| Related Protocol | EQR (Energy Efficient QoS Routing Protocol) |

Gymkhana is distributed and it consists of two phases. The first phase is the *Gymkhana protocol,* which aims at finding all possible paths from source toward the destination [7]. It uses the AODV protocol for the forwarding process. The second phased is the Gymkhana algorithm used by the destination, which evaluates the connectivity of paths by using the Laplacian matrix.

GYMKHANA PROTOCOL: Gymkhana protocol is used to collect the path level view of all possible paths from a source towards a destination by using an AODV – style mechanism. A source node S broadcasts a route request (RREQ) packet to discover all the possible paths towards the destination, D. RREQ packet passes among 'k' possible paths.

The RREQ arrives at the D at the *"l"* path where $l = 1,....,L$ contains two lists of elements, based on the number of hops, H [7]. First list, $L^a$, contains the Identification of the nodes that encounters in the path, *l*. Second list, $L^b$, contains the influence vectors of nodes encountered during the path. Both the lists are composed hop by hop. The first list is initialized with the identifier of node S while the second list is initialized with the influence vector, $I^s$

The intermediate node receiving the RREQ, checks whether its ID is already contained in $L^a$. If it is present in the list, then it discards the packet, else it adds its ID in the $L^a$ and its influence vector in $L^b$ and finally it broadcasts the RREQ. Therefore, the destination node D receives multiple RREQs packets and it has to select one path, based on the connectivity and by running the Gymkhana Algorithm. The selected path is informed to the source by sending Route Reply (RREP) packets in the reverse path [7]

GHYKHANA ALGORITHM: Gymkhana Algorithm is used to evaluate, the degree connectivity of the virtual graph levels, at the destination, by using the average cognitive Laplacian. It processes the contents of the received RREQs at a destination in two steps. They are:

*Step 1: Formation of a virtual graph for each path(s).*

A virtual graph, associated with an each path is defined as a virtualization of the chain representing the path and the node in such graph is called "virtual node". The goal of this step is to have a representation of the path in terms of nodes composing the paths and PUs affecting these nodes. Between two virtual nodes there exists an edge only of these two nodes are consecutive nodes in the same path.

They are two forms of edges, i.e horizontal and vertical edges. A horizontal edge indicates the node receives a packet from another node of the same channel. In vertical edge, a node receives a packet from another node which is from a different channel. Each edge in the virtual graph has an associated weight received from two different factors. They are: the activity of PUs and the cost of switching of channels.

An example to show how virtual graphs are formed is given in the following Figure 7.

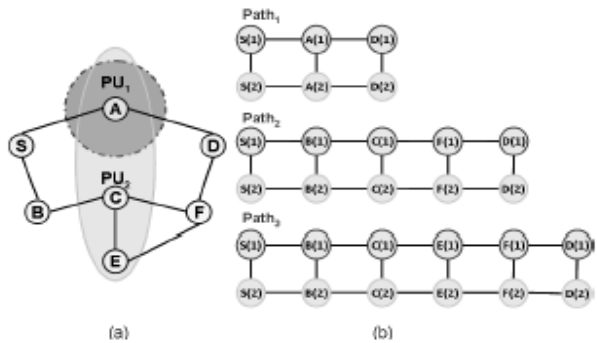

Figure 7 (a) A network topology consists of two PUs.
(b) Virtual graphs for the path S to D

From the Figure 7(a), it is shown that there is a network topology with two primary users and seven secondary users. There are three possible paths from S towards D. They are: S – A – D, S – B – C – F – D and  S – B – C – F – D. Figure 7(b) shows the three virtual graphs associated to these paths.

*Step 2: Laplacian of the virtual graph*

To build the Adjacency matrix, the link weights of each node can be used. Adjacency matrix will be assigned with the value zero if two nodes of the virtual graph are not connected by an edge; else the weight of the edge is assigned to it. It can be represented by $A(k)$. The degree of node is equal to the weights of edges related to that node, which is represented as $D(k)$. Therefore, the cognitive Laplacian matrix, $L(k) = D(k) - A(k)$.[7]

The metric used to evaluate the connectivity of a path is then based on the eigenvalues of $L(k)$. i.e by considering the second smallest eigenvalue of $L(k)$ is taken for the evaluation of the connectivity of the path. Some of the possible metric taken into account are: hop count, PUs activity, amount of SOPs switching along the path. The Table 5 summarizes the GYMKHANA protocol.

TABLE 5. SUMMARY OF GYMKHANA PROTOCOL

| Objective / Features | Algorithm used / Explanation |
|---|---|
| Goal | To propose a new routing scheme based on CRAHNs connectivity |
| Working of the protocol | Broadcast the packets using AODV protocol, represents the data in the form of a virtual graph and evaluate the cost by using the Laplacian matrix. |
| Techniques used | Gymkhana Algorithm |
| Path Discovery | All possible paths are found, which is less PU interference, using Virtual Graphs |
| Protocol nature | On Demand Routing Protocol |
| Best Path Selection | Based on the PUs activity, the cost of channel switching and the hop count |
| Routing Overhead | Yes. Secondary Users self – interference will increase the routing overhead |
| Evaluation is performed | By using Laplacian Graph theory |
| Related Protocol | Enhanced version of AODV |

## 2.6 Based on Clustering

Here, a distributed and efficient cluster-based spectrum and interference aware routing protocol is proposed, which incorporates the spectrum availability cost and interference metrics into the routing algorithm to find better routes.

### 2.6.1 UNITED NODE Protocol

For this protocol, a mobile CR ad hoc network environment with a number of primary and secondary nodes, where all nodes communicate with each other in their own networks, is considered. There is no communication (i.e. no cooperation) between primary and secondary networks. A novel algorithm, united nodes (UNITED) [17], is proposed for maximizing the network throughput and minimizing the end-to-end delay. The UNITED operates autonomously in a distributed manner at every node.

Initially, the nodes organise themselves into several clusters by the clustering algorithm that is based on location, communication efficiency, network connectivity and spectrum availability. After the completion of cluster formation, routing is done according to the spectrum usage and interference metrics. Clusters adapt themselves dynamically with respect to spectrum availability, and the high mobility of the nodes.

The proposed clustering algorithm for mobile CR ad hoc networks makes autonomous decisions in a distributed manner. It is based on a combined weight metric that takes into account several system parameters such as distance, transmission power, mobility, remaining power of nodes and sensed information about available spectrum.[17]

Depending on both the application and the environment, the contribution of these parameters to the final metric value may vary. The set of nodes are partitioned into clusters and each node is allowed to join only a single cluster. The CH selection procedure is invoked at the time of the system activation for all nodes. Initially, each node constitutes a cluster itself. Whenever a node receives data from a new neighbor node, it starts establishing a new cluster.

Depending on the metrics, either a new cluster is established by merging or the original status is kept. Once a new cluster is established, further cluster merge operations are controlled by the CH. The number of nodes $d_n$ within a single cluster could not exceed a preset value of δ. For details, the CH selection algorithm is given in Figure 8.

```
Algorithm 1
1  begin
2      repeat
3          ∀ node n, compute :
4              the node degree Δ_n = |d_n − δ|;
5              the mobility measure M_n(t);
6              spectrum availability Sp_n(t);
7              weighted value W_n = αΔ_n + βM_n(t) + γSp_n(t);
8          choose the node with the highest weighted value as the Cluster Head;
9          if n_j is CH for n_i and n_k is CH for n_j then
10             ⌊ n_i reselects CH excluding n_j;
11         remove neighbour nodes of the chosen CH from set of nodes
12     until G = ∅;
```

Figure 8. CH Selection

A node goes into an unclustered state, if all of its links to other nodes within the cluster fail. Also, all nodes within the range of the primary transmission activity return to the unclustered state whenever a primary user activity is detected. When a node falls into an unclustered state, it runs cluster formation process which is described in Figure 9. Nodes in the network have to keep local information about the neighboring nodes such as ID, speed, location, direction, cluster size and cluster membership. Such records have to be time-stamped in order to be expired after a predetermined time threshold, Δti. A node that falls into an unclustered state because of primary user activity usually has valid neighbor information in its table. If the set of valid neighbors is not empty, the algorithm starts checking for these nodes to join a cluster. Otherwise, the algorithm proceeds with searching for new neighborhood nodes. The set of one-hop neighbors, S, is produced by collecting replies from CHs to periodic HELLO packets.

Nodes receiving the HELLO packet utilises the node ID, the spectrum and the mobility information, that is, SOPs, location, speed and direction of travel, included in the packet to decide whether to reply or to ignore. Packets, either originating from the nodes that are moving away or having little spectrum access opportunity because of heavy primary user activity in the area, are ignored[17]. Otherwise, the receiving node checks whether the maximum number of connections δ is exceeded. If not, it responds with a unicast response RESP packet. Upon receiving the first RESP packet, the unclustered node sets a timer to wait for all responses, that is, to let all neighboring nodes to have an opportunity and respond.

```
Algorithm 2
  Input: S;
  Output: id of CH;
1 begin
2   NewList = false;
3   while !NewList do
4     if S = ∅ then
5       NewList = true;
6       repeat
7         send hello packet with reply request;
8       until S ≠ ∅;
9     order(S);
10    while S ≠ ∅ do
11      i ← max(∀j ∈ S);
12      send Join_Request(i);
13      if Join_Response received then
14        if connection accepted then
15          return clustered(i);
16        else
17          S ← S − {i};
18    return clustered(self);
```

Figure 9. Cluster Formation

After an unclustered node produces the set of its one-hop neighbor CHs, the set is sorted with respect to their weighted metrics in descending order and a search for actual connection is started. The unclustered node starts with sending a join request JOIN_REQ packet to the first CH in the list. This process is repeated for each node in the sorted set of S in order until a cluster is formed or S = Ø. If cluster is not formed the node seeks for a new set of S. If such a set cannot be established, the unclustered node creates its own cluster by declaring itself a CH and terminates the algorithm.

TABLE 6. SUMMARY OF UNITED NODES PROTOCOL

| Objective / Features | Algorithm used / Explanation |
|---|---|
| Goal | To provide adaptability to the environment and to increase the throughput, by reducing the data delivery latency. |
| Working of the protocol | Broadcast the Hello packets to the neighbouring nodes. Response will be sent by using the RESP packet and the request for joining will be done using the Join_Req packet. |
| Techniques used | Clustering Algorithm |
| Path Discovery | All possible routes are found, where the PU activity are less. |
| Protocol nature | Formation of Cluster based On Demand by the SU |
| Best Path Selection | Based on the PUs activity, the cost spectrum availability and interference metrics. |
| Unique Feature | A route preservation method is used to repair the route because of PUs interference. |
| Related Protocol | SEARCH Protocol [10] |

## 3. CONCLUSIONS

Routing in MCRAHNs is very challenging due to spectrum unavailability, PUs movements, etc. In this paper, the foundations of Cognitive radio ad hoc networks are presented, followed by its classification. The Classification of MCRAHNs is classified into five types: Multi – channel usage, Link Modelling, Geographic routing, Spectrum awareness, Connectivity, and Clustering depending on its operation. The protocols are highlighted with their working; techniques used and finally with the summary of the protocols are presented. Though most of the protocols of Mobile Cognitive Radio Ad hoc networks are found common in wireless networks, there is a need to design new metrics to show the uniqueness of Cognitive radio ad hoc networks.

## REFERENCES

[1] C. K. Toh, "Ad Hoc Mobile Wireless Networks", Pearson, 2006.

[2] Simon Haykin, "Cognitive radio: brain – empowered wireless communications", IEEE journal on Selected Areas in Communications, Vol.23 no 2, pp 201 – 220, 2005.

[3] A. Cagatay Talay, D.Turgay Altilar, "RACON: a routing protocol for mobile cognitive radio networks", CoRoNet '09, Proceedings of the 2009 ACM workshop on Cognitive radio networks, pp73 – 78, 2009.

[4] Amjad Ali, Muddesar Iqbal, Adeel Baig, Xingheng Wang, " Routing Techniques in Cognitive radio networks: a survey", International Journal of Wireless & Mobile Networks, Vol 3, No.3, pp 96 – 110, 2011.

[5] L. Hou, K.H Yeung, K.Y Wong, "A vision of energy – efficient routing for cognitive radio ad hoc networks", 6th International Symposium on Wireless and Pervasive Computing, pp 1 – 4, 2011.

[6] I.F Akyildiz, W.Y. Lee and K.R. Chowdhury, "CRAHNs: Cognitive radio ad hoc networks," Ad Hoc Networsk, Vol 7, no 5, pp 810 – 836, 2009.

[7] A. Abbagnale, F. Cuomo, "Gymkhana: A connectivity – based Routing Scheme for Cognitive Radio Ad Hoc Networks", Infocom IEEE conference on computer Communications workshops, pp. 1 – 5, 2010.

[8] Li Zong –shou, Zhu Qi, " A Novel Cross layer routing protocol for CR Ad Hoc Network", 6th International conference on Wireless Communications Networking and Mobile computing, pp 1 – 4, 2010.

[9] A. Cagatay Talay, D.Turgay Altilar, " ROPCORN: Routing protocol for cognitive radio ad hoc networks", International conference on Ultra Modern Telecommunications & workshops, pp 1 – 6, 2009.

[10] K.R. Chowdhury, M. Di Felice, "SEARCH : A routing protocol for mobile cognitive radio ad – hoc networks", SARNOFF'09, IEEE Sarnoff Symposium, pp 1 – 6, 2009.

[11] Angela Sara Cacciapuoti, Marcello Caleffi, Luigi Paura, "Reactive routing for mobile cognitive radio ad hoc networks", Ad Hoc Networks, Special Issue on Cognitive Radio Ad Hoc networks, Volume 10, issue 5, pp 803 – 815, 2012.

[12] S.M Kamruzzaman, Eunhee Kim, Dong Geun Jeong, " Spectrum and energy aware routing protocol for cognitive radio ad how networks", IEEE International Conference on Communications, pp 1-5, 2011.

[13] Mitola III, "Cognitive Radio: An Integrated Agent Architecture for Software Defined Radio," Ph.D. thesis, Dept. of Teleinformatics, KTH Royal Institute of Technology, 2000.

[14] http://www.ece.gatech.edu/

[15] C. Perkins, E. Belding – Royer, and S. Das, "Ad hoc On Demand Distance Vector (AODV) routing", RFC 3561, 2003.

[16] Perkins, Charles E. and Bhagwat, Pravin, "*Highly Dynamic Destination-Sequenced Distance-Vector Routing (DSDV) for Mobile Computers*", ACM SIGCOMM 94, pp 234 – 244, 1994.

[17] A.C Talay and D.T Altilar, "United Nodes: Cluster – based routing protocol for mobile cognitive radio networks", IET Communication 2011, Vol.5, Issue 15, pp. 2097 – 2105.

**Authors**

**Ms S. Selvakanmani** received the B.Tech degree in Information Technology from Velammal Engineering College, India in 2004 and the M.E degree in Digital Communication and Network Engineering from Arulmigu Kalasalingam College of Engineering, India, in 2006. Currently she is pursuing her doctoral degree under Anna University of Technology, India. She is currently working as Assistant Professor in the department of Computer Science and Engineering in Velammal Institute of Technology, India. Her research interest includes Computer networks, wireless communication, and computer algorithms.

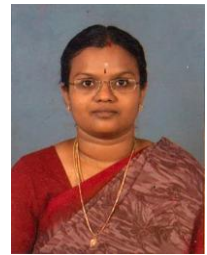

**Dr. M. Sumathi** received the B.E degree in Electronics and Communication Engineering from PSG College of Technology, India in 1992 and the M.E degree in Optical Communication from College of Engineering, Chennai, India in 2004. She pursued her doctoral degree under Anna University, Chennai, India in 2010. Currently she is working as Professor in the department of Electronics and Communication Engineering at Velammal Engineering College, Chennai, India. Her research interest includes WDM networks, optical communication, and computer networks. She is a member of OSA, IEEE and ISTE.

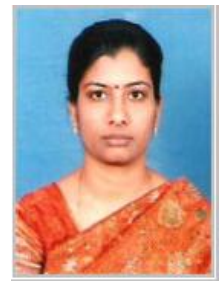